\documentclass[11pt]{article}
\usepackage{array}
\usepackage{graphicx}
\usepackage{cite}
\usepackage{textpos}

\title{Majorana and Condensed Matter Physics\footnote{Chapter for ``The  Physics of Majorana'', to be published by Cambridge University Press.}}
\author{Frank Wilczek}

\begin{document}

\maketitle

\begin{textblock*}{5cm}(11cm,-8.2cm)
\fbox{\footnotesize MIT-CTP-4542}
\end{textblock*}


\medskip

Ettore Majorana contributed several ideas that have had significant, lasting impact in condensed matter physics, broadly construed.   In this chapter I will discuss, from a modern perspective, four important topics that have deep roots in Majorana's work.       
\begin{enumerate}
\item{\it Spin Response and Universal Connection}:  In \cite{spinPaper}, Majorana considered the coupling of spins to magnetic fields.  The paper is brief, but it contains two ingenious ideas whose importance extends well beyond the immediate problem he treated, and should be part of every physicist's toolkit.   The first of those ideas, is that having solved the problem for spin $\frac{1}{2}$, one can deduce the solution for general spins by pure algebra.   Majorana's original construction uses a rather specialized mathematical apparatus.  Bloch and Rabi, in a classic paper \cite{blochRabi}, brought it close to the form discussed below.   Rephrased in modern terms, it is a realization -- the first, I think, in physics -- of the universality of non-abelian charge transport (Wilson lines).    
\item{\it Level Crossing and Generalized Laplace Transform}: In the same paper, Majorana used an elegant mathematical technique to solve the hard part of the spin dynamics, which occurs near level crossings.   This technique involves, at its center, a more general version of the Laplace transform than its usual use in constant-coefficient differential equations.   Among other things, it gives an independent and transparent derivation of the celebrated Landau-Zener formula for non-adiabatic transitions.  (Historically, Majorana's work on problem, and also that of Stueckelberg \cite{stueckelberg}, was essentially simultaneous with Landau's \cite{landau} and Zener's \cite{zener}.)  Majorana's method is smooth and capable of considerable generalization.  It continues to be relevant to contemporary problems.
\item{\it Majorana Fermions, From Neutrinos to Electrons}: Majorana's most famous paper \cite{fermionPaper} concerns the possibility of formulating a purely real version of the Dirac equation.  In a modern interpretation, this is the problem of formulating equations for the description of spin-$\frac{1}{2}$ particles that are their own antiparticles: Majorana fermions.   Majorana's original investigation was stimulated in part by issues of mathematical esthetics, and in part by the physical problem of describing the then hypothetical neutrino.  As described elsewhere in this volume \cite{particleChapter}, the immediate issue it raised -- i.e., whether neutrinos are their own antiparticles -- has become a central issue in modern particle theory, that still remains unresolved experimentally.   Here I will show that Majorana's basic idea is best viewed within a larger context, where we view it as a tool for analyzing unusual, number-violating forms of mass.  Within that larger conception we find, notably, that electrons within superconductors, when their energy is near to the fermi surface, behave as Majorana fermions.   
\item{\it Majorinos and Emergent Symmetry}:  A very special and unusual kind of Majorana fermion -- which, indeed, stretches the notion of ``fermion'' -- has been the focus of much recent attention, both theoretical and experimental.   These are zero energy excitations that are localized around specific points, such as the ends of wires, or nodes within circuits of specially designed superconducting materials.    Regarded as particles, they represent $0+1$ dimensional (0 space, 1 time) versions of the more familiar $3+1$ dimensional Majorana fermions, and in addition they have zero mass.   Since they are so small both in extent and in weight, I have proposed \cite{majorinoAlgebra} to call them {\it Majorinos}.   Majorinos have remarkable properties, including unusual quantum statistics, and might afford an important platform for quantum information processing \cite{kitaevWire}.   
\end{enumerate}

\section{Spins Response and Universal Connection}

The time-dependent Hamiltonian 
\begin{equation}\label{spinDynamics}
H(t) ~=~ \gamma \, {\vec B} (t) \cdot {\vec J}
\end{equation}
governs the interaction between a spin $\vec J$ and a time-dependent magnetic field $\vec B(t)$.   The dynamical evolution implied by Eqn.(\ref{spinDynamics}),
\begin{equation}\label{spinSchrodinger}
i \frac{d \psi}{d t} ~=~ H \psi
\end{equation}
for the spin wave-function $\psi$ is the most basic building-block for all the many applications of magnetic resonance physics.  Its solution \cite{spinPaper} is foundational.  

The $\vec J$ matrices obey the commutation relations of angular momentum.  Applied to a particle (which may be a nucleus, or an atom or molecule) of spin $j$, they obey the algebra 
\begin{eqnarray}
\left[ J_k, J_l \right] ~&=&~ i \, \epsilon_{klm} \, J_m \nonumber \\
{\vec J}\ ^2 ~&=&~ j(j+1)
\end{eqnarray}
which is realized in a $2j+1$ dimensional Hilbert space.  The representation is essentially unique, up to an overall unitary transformation.   

In the simplest case, $j=\frac{1}{2}$, we have a two-dimensional Hilbert space, we can take $\vec J = \frac{\vec \sigma}{2}$, with the Pauli matrices
\begin{eqnarray}
\sigma_1 ~&=&~ \left(\begin{array}{cc}0 & 1 \\1 & 0\end{array}\right) \nonumber \\
\sigma_2 ~&=&~ \left(\begin{array}{cc}0 & -i \\i & 0\end{array}\right) \nonumber \\
\sigma_3 ~&=&~ \left(\begin{array}{cc}1 & 0 \\0 & -1\end{array}\right) 
\end{eqnarray}
The differential equation Eqn.\,(\ref{spinSchrodinger}) becomes a system of two first-order equations, that is reasonably tractable.  (See, for example, the following Section!)

For larger values of $j$ we need to organize the calculation cunningly, lest it spin out of control.   As Majorana recognized, group theory comes to our rescue.  Here I will rephrase his basic insight in modern language.   

We can write the solution of Eqns.\,(\ref{spinDynamics},\ref{spinSchrodinger}) as a time-ordered exponential
\begin{equation}\label{wilsonLine}
\psi (t) ~=~ T \, \exp \, (-i \gamma \int\limits_0^t \, dt \vec B \cdot \vec J) \psi (0) ~\equiv~ U(t) \psi (0)
\end{equation}
Now if we call $B^j \rightarrow A^j_0$, and view it as the scalar potential associated with an $SU(2)$ symmetry, we recognize that the ordered exponential in Eqn.\,(\ref{wilsonLine}) is an example of a {\it Wilson line}, which implements parallel transport according to the nonabelian potential $A$.  In the context of gauge theories, it is a familiar and basic result that if one knows how to parallel transport the basic representation (vectors), then one can construct the parallel transport of more complicated representations (tensors) uniquely, by algebraic manipulations, without solving any additional differential equations.  This is a consequence -- in some sense it is the essence -- of the universality of nonabelian gauge couplings.      

In the present case, we can carry the procedure out quite explicitly.   The most convenient way is to realize the spin-$j$ object as a {\it symmetric spinor} \cite{symmetricSpinor}.  Thus we write
\begin{equation}
\psi ~=~ \psi^{\alpha_1 ... \alpha_{2j}}
\end{equation}
where the $\alpha_k$ are two-valued indices, and $\psi$ is invariant under their permutation.   Then if $U^{(\frac{1}{2})} = V^\alpha_\beta$ is the Wilson line for spin $\frac{1}{2}$, the Wilson line for spin $j$ will be
\begin{equation}\label{mainResult}
U^{(j)} ~=~ (U^{(j)})^{\alpha_1 ... \alpha_{2j}}_{\beta_1 ... \beta_{2j}}  ~=~ V^{\alpha_1}_{\beta_1} ... V^{\alpha_{2j}}_{\beta_{2j}} ~\equiv~ V\, \otimes ... \otimes \, V
\end{equation}
where the tensor product contains $2j$ copies.
This follows from the underlying realization of the algebra of $\vec J^{(j)}$ for spin $j$ as 
\begin{equation}
\vec J^{(j)} ~=~ \vec J^{(\frac{1}{2})} \otimes 1 \otimes ... \otimes 1 \, + \, ... \, + \, 1 \otimes ... \otimes 1 \otimes \vec J^{(\frac{1}{2})}
\end{equation}
including $2j$ summands.
From this point on, starting from Eqn.\,(\ref{mainResult}), the algebra required to develop concrete formulae, ready for use, is straightforward.  For Majorana to have exploited nonabelian symmetry in such a sophisticated way, already in 1932, was not only useful, but visionary.

\section{Level Crossing and Generalized Laplace Transform}

Having emphasized the value of solutions of Eqns.\,(\ref{spinDynamics},\ref{spinSchrodinger}) for spin-$\frac{1}{2}$, we turn to the the task of obtaining them.   The general problem can only be solved numerically.  There is, however, a powerful approximate result available when the evolution is slow and smooth, namely the adiabatic theorem.  It states, roughly speaking, that if the magnetic field varies slowly on the scale set by the (inverse) energy splittings $| \gamma B|^{-1}$, states initially occupying an energy eigenstate will remain within that energy eigenstate, as it evolves continuously -- there are no ``quantum jumps''.    (One can refine this further, to make statements about the phase of the wave functions.  This leads into interesting issues connected with the geometric or Berry phase, but I will not pursue that line here.)   The conditions for the adiabatic theorem will fail when levels energy levels approach one another, so it is important to have a bridging formula, to interpolate through such episodes.   

The discussion that follows is basically an unpacking of Majorana's extremely terse presentation in \cite{spinPaper}.   

As an interesting, experimentally relevant situation, which captures the essence of the problem, let us consider the Schr\"odinger equation describing a situation where the field-induced splitting between two nearby levels goes through a zero.  We suppose that during the crucial time, when the total splitting is small, the we can take the field to behave linearly.  This leads us to analyze the Schr\"odinger equation 
\begin{eqnarray}\label{basicCrossingSchrodinger}
i \dot \psi ~&=&~ H \psi \nonumber \\
\psi ~&\equiv&~ \left(\begin{array}{c}\alpha \\\beta\end{array}\right) \nonumber \\
H ~&=&~ - \Delta t \, \sigma_3 \, + \, \varepsilon \, \sigma_1 
\end{eqnarray}
In terms of components, we have
\begin{eqnarray}\label{schrodingerComponents}
i \dot \alpha ~&=&~ - \Delta t \, \alpha \, + \, \varepsilon \beta \nonumber \\
i \dot \beta ~&=&~  + \Delta t \, \beta \, + \, \varepsilon \alpha
\end{eqnarray}
By differentiating the first member of Eqn.\,(\ref{schrodingerComponents}), using the second member to eliminate $\dot \beta$ in terms of $\alpha, \beta$, and finally using the first member again to eliminate $\beta$ in favor of $\alpha, \dot \alpha$ we derive a second-order equation for $\alpha$ alone:
\begin{equation}\label{secondOrderAlpha}
i \ddot \alpha ~=~ i \Delta \alpha - \varepsilon^2 \alpha - (\Delta t)^2 \alpha
\end{equation}

In order to use the generalized Laplace transform in a simple way, it is desirable to have $t$ appear only linearly.   We can achieve this by introducing
\begin{equation}
\gamma ~\equiv~ e^{-i \Delta \frac{t^2}{2}} \, \alpha
\end{equation}
In this way we find 
\begin{equation}\label{gammaEquation}
\ddot \gamma \, + \, 2 i \Delta t \dot \gamma \, + \, \varepsilon^2 \gamma ~=~ 0 
\end{equation}
For later use, note that 
\begin{equation}\label{betaFromGamma}
\dot \gamma ~=~ -i \epsilon e^{- i \Delta \frac{t^2}{2}} \, \beta
\end{equation}
so that we can obtain both $\beta$, and of course $\alpha$, from $\gamma$.

Now we express $\gamma(t)$ in the form
\begin{equation}
\gamma(t) ~=~ \int \, ds \, e^{st} \, f(s)
\end{equation}
over an integration contour to be determined.   Operating formally, we thereby re-express Eqn.\,(\ref{gammaEquation}) as
\begin{equation}\label{transformPreliminary}
\int \, ds e^{st} (- i \Delta t s \, + \, s^2 + \varepsilon^2 ) f(s) ~=~ 0
\end{equation}
Using 
\begin{equation}
e^{st} \, st \, f (s) ~=~ \frac{d}{ds} (e^{st} sf) - e^{st} \frac{d}{ds} (s f)
\end{equation}
we cast Eqn.\,(\ref{transformPreliminary}) into the tractable form 
\begin{equation}\label{transformFinal}
\int \, ds e^{st} \Bigl(  \Delta s f^\prime  \, + \, ( \Delta  + i \, s^2 + i \varepsilon^2 ) f  \Bigr)  ~=~ 0
\end{equation}
under the assumption that $e^{st} sf$ vanishes at the ends of the contour (if any).   This leads to a simple differential equation for $f(s)$, which is solved by
\begin{equation}
f (s) ~=~ A s^{-1-  i \frac{\varepsilon^2}{2\Delta}} \, e^{-i \frac{s^2}{2\Delta}}
\end{equation}
for constant $A$, which we will choose to be $A=1$.  We will also assume $\Delta > 0$.

As $\tau \rightarrow \pm \infty$ the levels are highly split, and the adiabatic theorem comes into play, forbidding further transitions.   So the most interesting problem, to compute the probability for transitions which violate the adiabatic theorem -- ``quantum jumps'' --  can be investigated by studying the asymptotics as $\tau \rightarrow \pm \infty$.   In those limits our contour integral contains a parametrically large exponential, and we can try to exploit the stationary phase approximation.   The stationary phase points, in those limits, occur at
\begin{equation}
s ~=~ - i \Delta t
\end{equation}
A short calculation reveals that we should orient our contour with $s - s_0 \propto (1-i)$ near these points, to achieve steepest descent.   This suggests that we use the contours
\begin{equation}\label{candidateContours}
s ~=~ p (1-i) - i \Delta t
\end{equation}
with $p$ real.   One verifies that $e^{st} sf$ vanishes rapidly for large $| p |$, so these contours satisfy our consistency condition.   

There is a subtlety here that proves crucial. In defining $s^{-1 - i \frac{\varepsilon^2}{2 \Delta}} \equiv e^{\ln s  ( -1 - i \frac{\varepsilon^2}{2\Delta})} $ we must choose a branch of the logarithm.  Let us take its principal value, maintaining a cut along the negative real axis.   The candidate contours of Eqn.\,(\ref{candidateContours}) cross the real axis at
$s = p = -i \Delta \tau$.  We see -- noting $\Delta > 0$ -- that for negative $t$ the contour avoids the cut, but for positive $t$ we must deform the contour to avoid the cut.  We do this by allowing it to run above just above the cut from $s = i \epsilon - i \Delta \tau$ to $s = i \epsilon$ and the back below the cut from $s = -i \epsilon$ to $s = -i \epsilon - i \Delta \tau$, where it rejoins its original trajectory.  

For $\tau < 0$ (and $| \tau |$ large) the saddle point saturates the integration, and we find
\begin{equation}\label{saddlePointResult}
\gamma ~\approx~ \sqrt{\pi \Delta} \ \cdot e^{- i \frac{\Delta t^2}{2} } \ \cdot (\Delta | t | )^{-1 - i\frac{\varepsilon^2}{2\Delta} } \ \cdot e^{-i\frac{\pi}{2} } e^{\frac{\pi \varepsilon^2}{4 \Delta}}
\end{equation}
Let us review the origin of the various factors.   The first arises from the final Gaussian integral over $p$.  The second is the value of the exponential at the stationary phase point.   The third and fourth come, respectively, from inserting the two terms of the logarithm 
\begin{equation}\label{complexLog}
\ln (- i \Delta t ) ~=~ \ln (i \Delta |t | ) ~=~ \ln (\Delta |t| ) + i \frac{\pi}{2}
\end{equation}
into 
\begin{equation}
s^{-1-  i \frac{\varepsilon^2}{2\Delta}} ~=~ e^{\ln s (-1-  i \frac{\varepsilon^2}{2\Delta}) }
\end{equation}
We see that $\gamma$ vanishes for large $-t $ while, according to Eqn.(\ref{betaFromGamma}), $\beta$ has a finite limit.   

For $\tau > 0$ we still have a saddle point contribution, very similar to Eqn.\,(\ref{saddlePointResult}), but with the exponential factor $e^{\frac{\pi \varepsilon^2}{4 \Delta}} \rightarrow e^{-\frac{\pi \varepsilon^2}{4 \Delta}}$.  The difference arise from the change in sign of the imaginary part of the complex logarithm in Eqn.\,(\ref{complexLog}) We will also have a contribution from the integral over the cut.  It is not impossible to evaluate that integral, but we can get the most important result without it.   We need only to observe that the contour integral is non-oscillatory, so it doesn't contribute to $\beta$.  
Thus in the limit $\tau \rightarrow - \infty$ the wave function contains only a lower ($\beta$) term, and is proportional to 
$e^{\pi\frac{ \varepsilon^2}{4 \Delta}}$, while in the limit $\tau \rightarrow - \infty$ the $\beta$ term is proportional to 
$e^{-\pi \frac{\varepsilon^2}{4 \Delta}}$, with the other factors either in common or pure phases.   We conclude that the probability for the state to evolve from negative to positive energy, making a quantum jump, is the square of the ratio, or
\begin{equation}
e^{-\pi\frac{\varepsilon^2}{\Delta}}
\end{equation}
-- which is the classic ``Landau-Zener'' result.   

Majorana's method is very well adapted to generalizations involving multiple level crossings.   By keeping track of the phases, one can also find useful analogues of the geometric phase; and by applying group theory, also analyze {\it multiplet\/} crossings systematically \cite{crossing}.   It is interesting to contemplate an ambitious program, where one might infer the time-dependence of states from the time-dependent energy levels to, by patching together integral representations of the type we've just discussed, using these sorts of linear models to interpolate between adiabatic evolution.

\section{Majorana Fermions and Majorana Mass: From Neutrinos to Electrons}

\subsection{Majorana's Equation}

In 1928 Dirac \cite{dirac} proposed his relativistic wave equation for electrons, today of course known as the Dirac equation.    This was a watershed event in theoretical physics, leading to a new understanding of spin, predicting the existence of antimatter, and impelling -- for its adequate interpretation -- the creation of quantum field theory.   It also inaugurated a new method in theoretical physics, emphasizing mathematical esthetics as a source of inspiration.    Indeed, when we venture into the depths of the quantum world, ``physical intuition'' derived from the common experience of humans interacting with the world of macroscopic objects is of dubious value, and it seems inevitable that other, more abstract principles must take its place to guide us.   Majorana's most influential work \cite{fermionPaper} builds directly on Dirac's, both in content and in method.  By posing, and answering, a simple but profound question about Dirac's equation, Majorana expanded our concept of what a particle might be, and, on deeper analysis, of what mass is.   
For many years Majorana's ideas appeared to be ingenious but unfulfilled speculations; recently, however, they have come into its own.   They now occupy a central place in several of the most vibrant frontiers of modern physics including -- perhaps surprisingly -- condensed matter physics.

To appreciate the continuing influence of Majorana's famous ``Majorana fermion'' paper \cite{fermionPaper} in contemporary condensed matter physics, it will be helpful first to review briefly its central idea within its original context of particle physics.  I will do this in modern language, but with an unusual emphasis that leads naturally into the generalizations we'll be considering.  The companion chapter \cite{particleChapter} contains a much more extensive discussion of the particle physics around Majorana fermions.

Dirac's equation connects the four components of a field $\psi$.  In modern (covariant) notation it reads
\begin{equation}\label{diracEquation}
(i\gamma^\mu \partial_\mu - m ) \psi \ = \ 0
\end{equation}
The $\gamma$ matrices must obey the rules of {\it Clifford algebra}, i.e. 
\begin{equation}\label{cliffordAlgebra}
\{ \gamma^\mu \gamma^\nu \}
 \ \equiv \ \gamma^\mu \gamma^\nu \ + \ \gamma^\nu \gamma^\mu
 \ = \ 2 \eta^{\mu \nu} 
 \end{equation}
where $\eta^{\mu \nu}$ is the metric tensor of flat space.  Spelling it out, we have
\begin{eqnarray}
(\gamma^0 )^2 \ &=& \ - (\gamma^1 )^2 \ = \ -(\gamma^2 )^2  \ = \ -(\gamma^3 )^2 \ = \ 1; \\
\gamma^j \gamma^k \ &=& \ - \gamma^k \gamma^j \ \ {\rm for} \ \ i \neq j
\end{eqnarray}
(I have adopted units such that $\hbar = c = 1$.) Furthermore we require that $\gamma^0$ be Hermitean, the others anti-Hermitean.  These conditions insure that the equation properly describes the wave function of a spin-$\frac{1}{2}$ particle with mass $m$.   

Dirac found a suitable set of $4\times 4$ $\gamma$ matrices, whose entries contain both real and imaginary numbers.    For the equation to make sense, then, $\psi$ must be a complex field.   Dirac, and most other physicists, regarded this as a good feature, because electrons are electrically charged, and the description of charged particles requires complex fields, even at the level of the Schr\"odinger equation.   Another perspective comes from quantum field theory. In quantum field theory, if a given field $\phi$ creates the particle $A$ (and destroys its antiparticle $\bar A$), then the complex conjugate $\phi^*$ will create $\bar A$ and destroy $A$.   Particles that are their own antiparticles must be associated with fields obeying $\phi = \phi^*$, that is, real fields.   The equations for particles with spin 0 (Klein-Gordon equation), spin 1 (Maxwell equations) and spin 2 (Einstein equations, derived from general relativity) readily accommodate real fields, since the equations are formulated using real numbers.  Neutral $\pi$ mesons $\pi^0$, photons, and gravitons are their own antiparticles, with spins $0, 1, 2$ respectively.  But since electrons and positrons are distinct, the associated fields $\psi$ and $\psi^*$ must be distinct; and this feature appeared to be a natural consequence of Dirac's equation.  

In his 1937 paper Majorana posed, and answered, the question of whether equations for spin-$\frac{1}{2}$ fields must {\it necessarily}, like Dirac's original equation, involve complex numbers. Considerations of mathematical elegance and symmetry both motivated and guided his investigation. Majorana discovered that, to the contrary, there is a simple, clever modification of Dirac's equation that involves only real numbers.  Indeed, once having posed the problem, it is not overly difficult to find a solution.   
The matrices
\begin{eqnarray}
 \tilde\gamma^0 ~&=&~ \sigma_2 \otimes \sigma_1 \nonumber  \\
 \tilde\gamma^1 ~&=&~ i\sigma_1 \otimes 1  \nonumber \\
 \tilde\gamma^2 ~&=&~ i\sigma_3 \otimes 1 \nonumber \\
 \tilde\gamma^3 ~&=&~ i\sigma_2 \otimes \sigma_2
\end{eqnarray}
or, in expanded form:
\begin{eqnarray}
\tilde\gamma^0  ~&=&~
\left(\begin{array}{cccc} \ 0 &\  0 & \ 0 & -i \\ \ 0 &\  0 & -i & \ 0 \\ \ 0 &\  i & \ 0 & \ 0 \\ \ i & \ 0 & \ 0 & \ 0\end{array}\right) \nonumber \\
\tilde\gamma^1 ~&=&~
\left(\begin{array}{cccc} \ 0 & \ 0 & \ i & \ 0 \\ \ 0 & \ 0 & \ 0 &\  i \\ \ i & \ 0 & \ 0 & \ 0 \\ \ 0 & \ i &\  0 & \ 0\end{array}\right) \nonumber \\
\tilde\gamma^2 ~&=&~
\left(\begin{array}{cccc} \ i &\ 0 & \ 0 & \ 0 \\ \ 0 & \ i &\  0 & \ 0 \\  \ 0 &\  0 & -i & \ 0 \\ \ 0 & \ 0 & \ 0 & -i \end{array}\right) \nonumber \\
\tilde\gamma^3 ~&=&~
\left(\begin{array}{cccc} \ 0 & \ 0 & \ 0 & -i \\ \ 0 & \ 0 &\ i & \ 0 \\ \ 0 & \ i & \ 0 & \ 0 \\-i & \ 0 & \ 0 & \ 0\end{array}\right)
\end{eqnarray}
satisfy the same algebra Eqn.\,(\ref{cliffordAlgebra}) as Dirac's, and are purely imaginary.  Majorana's version of the Dirac equation
\begin{equation}
(i\tilde\gamma^\mu \partial_\mu - m ) \tilde \psi \ = \ 0
\end{equation}
therefore has the same desirable symmetry properties, including Lorentz invariance.  But
since the $\tilde \gamma^\mu$ are pure imaginary the $i \tilde \gamma^\mu$ are real, and so Majorana's equation can govern a real field $\tilde \psi$.

With this discovery, Majorana made the idea that spin-$\frac{1}{2}$ particles could be their own antiparticles theoretically respectable, that is, consistent with the general principles of relativity and quantum theory. In his honor, we call such hypothetical particles Majorana fermions.  

\subsection{Analysis of Majorana Neutrinos}

Majorana speculated that his equation might apply to neutrinos. In 1937, neutrinos were themselves hypothetical, and their properties unknown. The experimental study of neutrinos commenced with their discovery \cite{reines} in 1956, but their observed properties seemed to disfavor Majorana's idea. Specifically, there seemed to be a strict distinction between neutrinos and antineutrinos.   In recent years, however, Majorana's question has come back to life.  

The turning point came with the discovery of neutrino oscillations \cite{neutrinoOscillations}.  Neutrino oscillations provide evidence for mass terms.  Indeed, it is non-diagonal mass terms, connecting neutrinos with different lepton numbers, that cause freely propagating neutrinos to mix.    Mass terms, diagonal or not, are incompatible with chiral projections.  Thus the familiar ``left-handed neutrino'', which particle physicists worked with for decades, can only be an approximation to reality.   The physical neutrino must have some admixture of right-handed chirality.  

Thereby a fundamental question arises: Are the right-handed components of neutrinos something entirely new -- or could they involve the same degrees of freedom we met before, in antineutrinos?  (Usually these questions are phrased in the historically appropriate but cryptic form: Are neutrinos Majorana particles?)  At first hearing that question might sound quite strange, since neutrinos and anti-neutrinos have quite different properties.   How could there be a piece of the neutrino, that acts like an antineutrino?   But of course if the size of the unexpected piece is small enough, it can be compatible with observations.  Quantitatively: If the energy of our neutrinos is large compared to their mass, the admixture of opposite chirality will be proportional to $m/E$.   To explain the phenomenology of neutrino oscillations, and taking into account cosmological constraints, we are led to masses $m < {\rm eV}$, and so in most practical experiments $m/E$ is a very small parameter.  These considerations raise the possibility that neutrinos and antineutrinos the same particles, just observed in different states of motion.  The observed distinctions might just represent unusual spin-dependent (or, more properly, helicity-dependent) interactions.


To pose the issues mathematically, we must  describe a massive spin-$\frac{1}{2}$ particle using just two -- not four -- degrees of freedom. We want the antiparticle to involve the same degrees of freedom as the particle. Concretely, we want to investigate how the hypothesis
\begin{equation}\label{MajoranaHypothesis}
\psi_R ~\stackrel{?}{=}~ \psi_L^{\ *}
\end{equation}
might be compatible with non-zero mass.  Eqn.\,(\ref{MajoranaHypothesis}) embodies, in precise mathematical form, the idea that antineutrinos are simply neutrinos in a different state of motion, i.e. with different helicity.  

If $\psi$ is a real field, described by Majorana's version of the Dirac equation, then 
\begin{equation}
(\psi_L)^{\ *} ~\equiv~ (\frac{1 - \gamma_5}{2} )^{\ *} \psi ~=~ (\frac{1 + \gamma_5}{2} ) \psi ~\equiv~ \psi_R
\end{equation}
since $\gamma_5 \, \equiv \, i \gamma^0 \gamma^1 \gamma^2 \gamma^3$ is pure imaginary.  Conversely, if
Eqn.\,(\ref{MajoranaHypothesis}) holds, we can derive both $\psi_L$ and $\psi_R$ by projection from a single four-component {\it real\/} field, i.e.  
\begin{equation}
\psi ~\equiv~ \psi_L + \psi_R ~=~ \psi_L + \psi_L^{\ *}
\end{equation}
This is the link between Majorana's mathematics and modern neutrino physics.  

\subsection{Majorana Mass}

With that background and inspiration, we can distill the essential novelty in the Majorana equation, which is a bit more subtle than is commonly stated.   What is distinctive is not merely the use of real fields.  After all, a complex field can always be written in terms of two real ones, as $\psi \, = \, {\rm Re} \psi + i {\rm Im} \psi$, and so any system of equations involving the complex field $\psi$ can be written as a larger system of equations involving only the real fields ${\rm Re} \psi$ and ${\rm Im} \psi$.    Rather, what is distinctive is the possibility of passing, in the description of a massive spin $\frac{1}{2}$ particle, from a Dirac field with eight real degrees of freedom (four complex components) to a Majorana field with four real degrees of freedom.  As we shall see, this requires an unusual, symmetry-breaking form of mass: Majorana mass.

\subsubsection{Broken Symmetry Aspect}

To highlight the innovation this requires, let us consider the chiral projection of the Majorana equation.  In general, applying a chiral projection to the Dirac equation gives us
\begin{equation}
i \gamma^\mu \partial_\mu \psi_L + M \psi_R ~=~ 0
\end{equation}
Both $\psi_L$ and $\psi_R$ naturally contain two complex components.   But in the Majorana specialization, $\psi_R$ is not independent, for it satisfies Eqn.\,(\ref{MajoranaHypothesis}).   We have, therefore,
\begin{equation}\label{MajoranaEquation}
i \gamma^\mu \partial_\mu \psi_L + M \psi_L^{\ *} ~=~ 0
\end{equation}
In this formulation, we see that the mass term is of an usual form.  It involves complex conjugating the field, and thus reads differently -- by a minus sign -- for its real and imaginary components.   It is naturally associated with breaking of the phase symmetry
\begin{equation}
\psi_L  ~\rightarrow~ e^{i\lambda} \psi_L
\end{equation}
or, of course, the corresponding number symmetry. 

\subsubsection{Formal Aspect: Grassmann Variables}

The appearance of Eqn.\,(\ref{MajoranaEquation}) is unusual, and we may wonder, as Majorana did, how it could arise as a field equation, following the usual procedure of varying a Lagrangian density.  That consideration led Majorana to another major insight.   The unprojected mass therm
\begin{equation}
{\cal L}_M ~\propto~ \bar \psi \psi ~=~ \psi^\dagger \gamma_0 \psi
\end{equation}
becomes, if we write everything in terms of $\psi_L$ (using Eqn.\,(\ref{MajoranaHypothesis}) 
\begin{equation}\label{MajoranaMassTerm}
{\cal L}_M ~\propto~ \psi^\dagger \gamma_0 \psi ~\rightarrow~ (\psi_L)^T \gamma_0 \psi_L + {(\psi_L^{\ *})}^T \gamma_0 \psi_L^{\ *}
\end{equation}
where $^T$ denotes transpose.  

In verifying that these terms are non-trivial, whereas the remaining cross-terms vanish, it is important to note that $\gamma_5$ is antisymmetric, i.e., that it changes sign under transpose.   That is true because $\gamma_5$ is both Hermitean and pure imaginary.   Thus we have, for example,  
\begin{eqnarray}
(\psi_L)^T \gamma_0 \psi_R ~&=&~ (\psi_L )^T \,  (\frac{1 - \gamma_5}{2})^T \, \gamma_0 \,  (\frac{1 + \gamma_5}{2}) \, \psi_R \nonumber \\
~&=&~ (\psi_L )^T\,  (\frac{1 +  \gamma_5}{2}) \, \gamma_0 \, (\frac{1 + \gamma_5}{2}) \,  \psi_R \nonumber \\
~&=&~ (\psi_L )^T\,  \gamma_0 \, (\frac{1 -  \gamma_5}{2}) \,  (\frac{1 + \gamma_5}{2}) \, \psi_R \nonumber \\
~&=&~ 0
\end{eqnarray}
(If we do not adopt Majorana's hypothesis Eqn.\,(\ref{MajoranaHypothesis}), the mass term takes the form
\begin{equation}
{\cal L}_{\rm conventional} ~\propto~ (\psi_R^{\, *})^T \psi_L + (\psi_L^{\, *})^T \psi_R
\end{equation}
which supports number symmetry $\psi_{L,R}  \rightarrow e^{i\lambda} \psi_{L,R}$.)

The survival of the remaining terms in Eqn.\,(\ref{MajoranaMassTerm}) is, as Majorana noted, also non-trivial.  In components, we have
\begin{equation}
(\psi_L)^T \gamma_0 \psi_L ~=~ (\gamma_0)^{jk}  (\psi_L)_j (\psi_L)_k
\end{equation}
Now $\gamma_0$, like $\gamma_5$, is antisymmetric (for the same reasons).   So in order for this term to survive, we must assume that the fields $\psi_j$ are {\it anticommuting\/} 
variables.  Majorana's bold invocation of such ``Grassmann numbers'', which have become ubiquitous in the modern field theory of fermions, was ahead of its time.   

With this understanding, variation of the mass term, together with the conventional kinetic term 
\begin{equation}
{\cal L} ~\propto~ {(\psi_L^{\ *})}^T \, \gamma_0 \,  i\gamma^\mu \partial_\mu \psi_L + h.c.
\end{equation}
will give us Eqn.\,(\ref{MajoranaEquation}).  

\subsubsection{Majorana Mass as Symmetry-Breaking Perturbation}

Eqn.\,(\ref{MajoranaMassTerm}) affords an instructive perspective on the Majorana mass term, which, we have argued, is a central innovation of \cite{fermionPaper}.   This perspective will be helpful in assessing the ``shocking'' result of our next section. 

By stripping away all kinematic details, we can define a faithful, transparent analogue of ``Majoranization'' through mass acquisition, for spin 0 bosons.  

Let $\phi$ and $\Delta$ be two complex boson fields, supporting a global $U(1)$ symmetry
\begin{equation}
(\phi, \Delta) ~\rightarrow~ (e^{i\alpha} \phi, e^{2i\alpha}\Delta)
\end{equation}
Our focus will be on how the properties of the $\phi$ quanta -- in particular, their masses -- change as $\Delta$ acquires a symmetry-breaking vacuum expectation value.   

The relevant Lagrangian is 
\begin{eqnarray}
{\cal L}_{\rm mass} ~&=&~ {\cal L}_{\rm conventional} \, + \, {\cal L}_{\rm Majorana} \nonumber \\
{\cal L}_{\rm conventional} ~&=&~ - M^2 \, \phi^* \phi \nonumber \\
{\cal L}_{\rm Majorana} ~&=&~ - \kappa \,  \Bigl( \Delta^* \phi^2 + \Delta (\phi^*)^2 \Bigr)
\end{eqnarray}
(taking, for convenience, $\kappa$ real).   

If the vacuum expectation value $\langle \Delta \rangle = 0$, then we have only the conventional mass term.  The quanta of $\phi$ are a degenerate pair -- particle and antiparticle -- with opposite charge but common mass $M$.  The states of definite charge are produced by $\phi$ and $\phi^*$.   

If the vacuum expectation value $\langle \Delta \rangle = v \neq 0$ (assumed, for convenience, real), then the quadratic terms in $\phi$ read
\begin{equation} 
{\cal L}_{\rm mass} ~=~ -M^2 \Bigl( ({\rm Re} \, \phi)^2 +  ({\rm Im} \, \phi)^2 \Bigr) \, - \, 2 \kappa v \Bigl( ({\rm Re} \, \phi)^2  -  ({\rm Im} \, \phi)^2 \Bigr)
\end{equation}
In this case, the quanta of definite mass are associated with the {\it real\/} fields ${\rm Re} \, \phi, {\rm Im} \, \phi$, and have masses 
\begin{eqnarray}
m_{\rm Re} ~&=&~ \sqrt {M^2 + 2 \kappa \Delta}  \nonumber \\
m_{\rm Im} ~&=&~ \sqrt {M^2 - 2 \kappa \Delta}
\end{eqnarray}

These particles with definite mass are ``Majorana'' particles, in the sense that they are their own antiparticles.  
On the other hand, if $\kappa \Delta$ is small, the practical effect of the splitting might be quite limited.  For example, let us suppose that other interactions (besides the mass) are more nearly diagonal in terms of $\phi$ and $\phi^*$, not ${\rm Re} \phi$ and ${\rm Im} \phi$, as will occur if the symmetry breaking is small.  Then the typical $\phi$ quantum produced in an interaction will evolve (in its rest frame) as
\begin{eqnarray}
| \phi (t = 0) \rangle ~&=&~ \frac{1}{\sqrt 2} ( | {\rm Re}\, \phi \rangle + i | {\rm Im} \, \phi \rangle )  \\
~&\rightarrow&~ \frac{1}{\sqrt 2} (e^{-i m_{\rm Re} t}  | {\rm Re}\, \phi \rangle + i e^{-i m_{\rm Im} t}  | {\rm Im} \, \phi \rangle ) \nonumber \\
~&=&~ e^{-i(m_{\rm Re} + m_{\rm Im} ) t/2} \Bigl( \cos \, \frac{m_{\rm Re} - m_{\rm Im}}{2} |\phi \rangle - \sin \frac{m_{\rm Re} - m_{\rm Im}}{2} | \phi^* \rangle \nonumber
\end{eqnarray}
in an evident notation.  The oscillation time 
\begin{equation}
(\frac{m_{\rm Re} - m_{\rm Im}}{2})^{-1} \ \stackrel{\kappa \Delta << M}{\approx} \  \frac{M}{\kappa \Delta}
\end{equation}
will be very long.  For shorter times, and considering the dominant interactions, it will be a good approximation to work with the ``non-eigenstates'' $| \phi \rangle, |\phi^* \rangle$, wherein the underlying Majorana structure is hidden.  

These considerations, which of course have their parallel for fermions, show that the concept ``Majorana particle'' should not be regarded as a binary, yes-or-no predicate.  For, as we have just seen, particles can be their own antiparticles and yet behave, for practical purposes (with arbitrary accuracy), as if they were not.   Rather the physically meaningful issues are the magnitude of Majorana mass terms, and the circumstances in which such mass term induce significant physical effects.

\subsection{Majorana Electrons}

After these preliminaries we are prepared to discuss the concept of Majorana electrons, which might otherwise sound absurd.

Electrons and antielectrons have opposite electric charge, and electric charge is most definitely an observable quantity, which might seem to preclude that electrons might be their own antiparticles.  
Inside superconductors, however, the situation is different.   Inside a superconductor, we have a condensate of Cooper pairs.   Heuristically:  A hole, in this environment, can be ``dressed'' by a Cooper pair, and come to look like a particle.  More formally: Inside a superconductor, the gauge symmetry associated to charge conservation is spontaneously broken, and electric charge is not a good quantum number, so one cannot invoke it to distinguish particles from holes.

To connect superconductivity with Majorana mass, consider how the formation of an electron pair condensate effects the ambient electrons.   From the electron-electron interaction term $\bar e \bar e ee$ -- suppressing spin indices, and considering only simple s-wave ordering -- we derive an effective interaction between electrons and the ambient condensate 
\begin{equation}\label{gapInteraction}
{\cal L}_{\rm electron-condensate} ~=~ \Delta^* \, e e \, + \, {\rm h.c.} ~\leftarrow~ \kappa \, \bar e \bar e e e + {\rm h.c.}
\end{equation}
that precisely mirrors the Majorana mass term!    


Famously, the interaction Eqn.\,(\ref{gapInteraction}) of electrons (and holes) with the condensate both mixes electrons with holes opens a gap in the electron spectrum at the fermi surface.   A close analogy between the opening of that gap and the generation of mass, by condensation, for relativistic fermions was already noted in Nambu's great work on spontaneously broken symmetry in relativistic particle physics \cite{nambu}.  Indeed, that analogy largely inspired his work.   We are emphasizing here, that in superconductors the mass in question is actually mass of Majorana's unusual, number-violating form.   

The induced Majorana mass, according to Eqn.\,(\ref{gapInteraction}), is $\sim 10 K = 10^{-3}$ eV, which is minuscule on the scale of the electron's intrinsic (normal) mass, and even small on the scale of ordinary Fermi energies.    It will dominate only for quasiparticles within a narrow range of the nominal Fermi surface.    Nevertheless, most of the phenomena of superconductivity follow from the existence of the gap and therefore, implicitly, from this Majorana mass.   

Very recently Beenakker has proposed a more pointed experimental demonstration of the Majorana nature of electrons in superconductors \cite{beenakker}.   While the details are complicated, the essential idea is that the Majorana mass term depends on the phase $e^{i\delta}$ of $\Delta$, and the relative phase of the particle and hole components of the Majorana quasiparticle will reflect that phase.   Thus if we bring together quasiparticles from two {\it different\/} superconductors, their overlap, and therefore their annihilation probability, will be proportional to $\cos^2 \frac{\delta_1 - \delta_2}{2}$.

\section{Majorinos}

As we have just discussed, inside superconductors the near-Fermi surface quasiparticle excitations have a Majorana character.  Another interesting theme in recent condensed matter physics is the importance of zero modes -- roughly speaking, mid-gap states -- in quasiparticle spectra.   They are typically associated with topological features, such as domain walls, vortices, or boundaries.  (Mathematical aside: The existence of these modes is connected to index theorems.)  

The conjunction of these two ideas leads us Majorana zero modes.   In a particle interpretation, the quanta associated to these modes are zero-mass Majorana particles.  Since they are localized on specific points, they can be considered as particles in $0+1$ (0 space, 1 time) dimension.   They are in many respects the most extreme Majorana (self-conjugate) particles, consistent with their origin as mid-gap states in superconductors, where the electron Majorana mass is most dominant.  They are not quite fermions, as we shall see, but obey an interesting generalization of Fermi statistics.  It will be convenient to have a name for such particles.  In view of their smallness both in mass and in spatial extension,  and their extreme Majorana character, the name {\it Majorino\/} suggests itself, and I shall adopt it here.

The existence of Majorana modes in condensed matter systems~\cite{Mourik2012,JackiwRossi, MooreRead,ReadGreen,AliceaRev,BeenakkerRev} is intrinsically interesting, in that it embodies a qualitatively new and deeply quantum mechanical phenomenon~\cite{Wilczek2009}. It is also possible that such modes might have useful applications, particularly in quantum information processing~\cite{Kitaev2006, NayakRMP}.  One feature that makes Majorana modes useful is that they generate a doubled spectrum.   When we have several Majorana modes, each produces an independent doubling.   Such repeated doubling generates a huge Hilbert space of degenerate states.   If we can control the dynamics of Majorinos, we can navigate through that Hilbert space.  That vision inspires research to enable quantum information processing using Majorinos. 

This section is structured as follows.  We first discuss the occurrence of Majorinos in a simple, discrete one dimensional model, following Kitaev's simple but profound analysis~\cite{kitaevWire}.  We then generalize to a more realistic model wire (still following Kitaev), and then to circuits.  In that context we identify a remarkable algebraic structure, that allows us to identify the Majorinos, to show that their existence is robust against interactions, and to exhibit the doubling phenomenon, in a transparent fashion.  Finally, we discuss a continuum field theory approach to Majorinos.   

\subsection{Kitaev Chain}

\subsubsection{Schematic Hamiltonian}

Let us briefly recall the simplest, yet representative, model for such modes, Kitaev's wire segment. We imagine $N$ ordered sites are available to our electrons, so we have creation and destruction operators $a_j^\dagger, a_k$, $1 \leq j, k \leq N$, with $\{ a_j , a_k \} \, = \{ a_j^\dagger , a_k^\dagger \} \, = 0$ and $\{ a_j^\dagger , a_k \} \, = \delta_{jk}$.   The same commutation relations can be expressed using the hermitean and antihermitean parts of the $a_j$, leading to a Clifford algebra, as follows:
\begin{eqnarray}\label{clifford}
\gamma_{2j-1} ~&=&~ a_j + a_j^\dagger \nonumber \\
\gamma_{2j} ~&=&~ \frac{a_j - a_j^\dagger}{i} \nonumber \\
\{ \gamma_k, \gamma_l \} ~&=&~ 2\, \delta_{kl}. 
\end{eqnarray}

Now let us compare the Hamiltonians
\begin{equation}\label{trivialH}
H_0 ~=~ -i \sum\limits_{j=1}^{N} \, \gamma_{2j -1} \gamma_{2j} 
\end{equation} 
\begin{equation}\label{majoranaModeH}
H_1 ~=~ -i \sum\limits_{j=1}^{N-1} \, \gamma_{2j} \gamma_{2j+1}.
\end{equation}
Since $-i \gamma_{2j-1}\gamma_{2j} = 2 a_j^\dagger a_j - 1$, $H_0$ simply measures the total occupancy.   It is a normal, if unusually trivial, electron Hamiltonian.  

$H_1$ strongly resembles $H_0$ but there are three major differences.
\begin{enumerate}   

\item
One difference emerges, if we re-express $H_1$ in terms of the $a_j$.   We find that it is local in terms of those variables, in the sense that only neighboring sites are connected, but that in addition to conventional hopping terms of the type $a_j a^\dagger_{j+1}$ we have terms of the type $a_j a_{j+1}$, and their hermitean conjugates.   The $aa$ type, which we may call superconductive hopping, does not conserve electron number, and is characteristic of a superconducting (pairing) state.    

\item A second difference grows out of a similarity: since the algebra Eqn.\,(\ref{clifford}) of the $\gamma_j$ is uniform in $j$, we can interpret the products $\gamma_{2j} \gamma_{2j+1}$ that appear in $H_1$ in the same fashion that we interpret the products $\gamma_{2j-1} \gamma_{2j}$ that appear in $H_0$, that is as occupancy numbers.   The effective fermions that appear in these numbers, however, are not the original electrons, but mixtures of electrons and holes on neighboring sites.  

\item The third and most profound difference is that the operators $\gamma_1, \gamma_{2N}$ do not appear at all in $H_1$.   These are the Majorana mode operators.   They commute with the Hamiltonian, square to the identity, and anticommute with each other.   The action of $\gamma_1$ and $\gamma_{2N}$ on the ground state implies a degeneracy of that state, and the corresponding modes have zero energy.   Kitaev \cite{kitaevWire} shows that similar behavior occurs for a family of Hamiltonians allowing continuous variation of microscopic parameters, i.e. for a universality class.  (We will reproduce the core of that analysis immediately below, and extend it in a different direction subsequently.)  Within that universality class one has hermitean operators $b_L, b_R$ on the two ends of the wire whose action is exponentially (in $N$) localized and commute with the Hamiltonian up to exponentially small corrections, that satisfy the characteristic relations $b_L^2 = b_R^2 = 1$.    In principle there is a correction Hamiltonian,
\begin{equation}
H_c ~\propto~ -i b_L b_R,
\end{equation}
that will encourage us to re-assemble $b_L, b_R$ into an effective fermion creation-destruction pair, and to realize $H_c$ as its occupation number, by inverting the construction of Eqn.\,(\ref{clifford}).   
But for a long wire and weak interactions we expect the coefficient of $H_c$ to be very small, since the modes excited by $b_L, b_R$ are spatially distant, and for most physical purposes it will be more appropriate to work with the local variables $b_L, b_R$, since these respond independently to spatially uncorrelated perturbations.  

\end{enumerate}

\subsubsection{Parametric Hamiltonian}

$H_1$ involved a very particular choice of parameters.  Here we show that can arise with a class of Hamiltonians of a physically plausible form, that contain continuous parameters, yet share its central qualitative feature -- that is, the emergence of Majorinos at the ends of the wire. 

Indeed, consider
\begin{equation}
H ~=~ \sum\limits_{j=1}^N  -w \, (a_j^\dagger a_{j+1} + a_{j+1}^\dagger a_{j} ) - \mu (a_j^\dagger a_{j} - \frac{1}{2} ) + \Delta a_j a_{j+1} + \Delta^* a^\dagger_{j+1} a^\dagger_{j}
\end{equation}
with the understanding $a^\dagger_{N+1} = a_{N+1} =0$. 
It describes a superconducting wire, with $N$ sites.   The first term, with coefficient $w$, is a conventional hopping term; the second term, proportional to $\mu$, is a chemical potential, and the remaining terms represent residual interactions with the superconducting gap parameter $\Delta \equiv e^{i\theta} |\Delta |$, here regarded as a given external field.   

The crucial issue, for us is the existence (or not) of zero energy modes.   So we look for solutions of the equation
\begin{equation}\label{0ModeEquation}
\left[ H, \sum\limits_{j=1}^{2N} c_j \gamma_j \right] ~=~ 0
\end{equation}
With the understanding that $\gamma_k \rightarrow 0$ when $k$ falls outside the allowed range $1\leq k \leq 2N$, we have, for $j$ even 
\begin{equation}
\left[ H, \gamma_j \right] ~\propto~ \mu  \gamma_{j-1}  + (w + |\Delta | ) \gamma_{j+1}  + (w - |\Delta | )  \gamma_{j-3}  \ \ \ \ \ \ j{\rm ~ even}
\end{equation}
while for $j$ odd
\begin{equation}
\left[ H, \gamma_j \right] ~\propto~ - \mu  \gamma_{j+1}  - (w + |\Delta | ) \gamma_{j-1}  - (w - |\Delta | )  \gamma_{j+3}  \ \ \ \ j{\rm ~ odd}
\end{equation}
Gathering terms proportional to $\gamma_l$ in Eqn. (\ref{0ModeEquation}), we have
\begin{eqnarray}\label{recurrenceRelation}
\mu c_{l+1}  + (w + |\Delta | ) c_{l-1} + (w - |\Delta | ) c_{l+3}  ~&=&~ 0 \ \ l {\rm  ~odd} \nonumber \\
-\mu c_{l-1}  - (w + |\Delta | ) c_{l+1} - (w - |\Delta | )  c_{l-3}  ~&=&~ 0 \ \ l {\rm  ~ even}
\end{eqnarray}
with the understanding that $c_j \rightarrow 0$ unless $1 \leq j \leq 2N$.  

Eqn. (\ref{recurrenceRelation}) is a three-term recurrence relation, that connects coefficients whose subscripts differ by two units.  At first neglecting boundary conditions, we construct candidate solutions in the form of power series, which contain only even or only odd terms.   
Putting $c_l = \alpha x^{(l +1)/2}$, for all odd $l$, we get a quadratic equation for $x$ (from the {\it second\/} line of Eqn. (\ref{recurrenceRelation}) ): 
\begin{equation}
x^{\rm odd}_\pm ~=~ \frac{- \mu \pm \sqrt{\mu^2 + 4 |\Delta|^2 - 4 w^2}}{2 (w + |\Delta | )}
\end{equation}
This leads us to look for zero modes in the form
\begin{equation}
b_1 ~=~ \sum\limits_j \  (\alpha_+ x_+^j  + \alpha_- x_-^j) \gamma_{2j-1}
\end{equation}
The zero-mode condition Eqn. (\ref{0ModeEquation}) is then satisfied, by construction, in bulk.   

Now we must attend to the boundaries.   In our formal manipulations we've ignored the prescriptions $c_l, \gamma_l \rightarrow 0$ for $l$ outside the physical range.   To have a legitimate solution at the ends, we must also impose
\begin{equation}\label{rightBoundaryCondition}
\alpha_+ x_+^{N+1} + \alpha_- x_-^{N+1} ~=~ 0
\end{equation}
and 
\begin{equation}\label{leftBoundaryCondition}
\alpha_+ x_+^{0} + \alpha_- x_-^{0} ~=~ \alpha_+  + \alpha_- ~=~ 0
\end{equation}

We reach different conclusions depending upon whether our candidate solutions are both growing (i.e., $|x_+ |, |x_- | > 1$), both shrinking ($|x_+ |, |x_- | < 1$) or we have one with each behavior.   If both are growing, then after imposing Eqn. (\ref{rightBoundaryCondition}), and normalizing the coefficient $\alpha_+ x_+^{N+1} =1$, we find 
\begin{equation}
\alpha_+  + \alpha_- = x_+^{-N-1} - x_-^{-N-1} \ ~\rightarrow_{N\rightarrow \infty}~ \  0
\end{equation}
so that Eqn. (\ref{leftBoundaryCondition}) is satisfied approximately, up to quantities that are exponentially small in $N$.   
For a half-infinite wire we can take the limit, and we have exact normalizable zero modes, concentrated on the (right-hand) boundary.   For large but finite $N$ we will need another small parameter in order to satisfy the second boundary condition.  Such a parameter is available: the energy.   So we will not get an exact zero-energy mode, but rather a normalizable mode with exponentially small energy, concentrated on the right-hand boundary.   

If both of our candidate solutions are shrinking, then a similar analysis reveals a normalizable near-zero energy mode concentrated on the left-hand boundary. Note that our earlier $H_1$ corresponds to $x_+ = x_- = 0$, so we have solutions concentrated on one site!  

If one candidate solution is growing, while the other is shrinking, the construction fails, for upon imposing one boundary condition we are forced into gross violations of the other.    

That completes our analysis of the candidate zero modes associated with linear combinations of $\gamma_{2l-1}$, i.e. concentrated on odd sites.   We can of course perform a parallel analysis of candidate zero modes concentrated on even sites.   Due to the re-sequencing of terms appearing in the two lines of Eqn. (\ref{recurrenceRelation}) the quadratic equation for the factor $x$ is reversed, so that it is solved instead by $x^{-1}$.  Thus candidate zero modes concentrated on even sites take the form
\begin{equation}
b_2 ~=~ \sum\limits_j \  (\beta_+ x_+^{-j}  + \beta_- x_-^{-j}) \, \gamma_{2j}
\end{equation}
We find that a normalizable near-zero mode of this even type occurs in precisely the same circumstances where we have a near-zero mode of the odd type; and that when both solutions exist, they are localized at opposite ends of the wire.

\subsection{Junctions and the Algebraic Genesis of Majorinos}

Now let us consider the situation where multiple Majorana modes come together to form a junction, as might occur in a network of superconducting wires of the sort analyzed above.  Several experimental groups are developing physical embodiments of Majorana modes, for eventual use in such quantum circuits.  (For a useful sampling of recent activity, see the collection of abstracts from the July 12-18 2013 Erice workshop: \cite{ericeWorkshop}.)  

A fundamental issue, in analyzing such circuits, is the behavior at junctions.   Pairs of Majorana modes can be organized into fermion creation and destruction operators, and their are allowed interaction terms which make that appropriate, but we can anticipate that general principles might be enough to give us one surviving Majorana mode at an {\it odd \/} junction. We will, in fact, identify a remarkably simple, explicit non-linear operator $\Gamma$ that can be considered the creator/destroyer of Majorinos.  $\Gamma$ obeys simple algebraic relations with the Hamiltonian, and implements a general doubling of the spectrum.   Its existence and properties are tightly connected to fermion number parity.  The emergent algebraic structure, in its power and simplicity, seems encouraging for further analysis and development of Majorana wire circuits.

The following considerations will appear more pointed if we
explain their origin in the following little puzzle.   Let us imagine we bring together the ends of three wires supporting Majorana modes $b_1, b_2, b_3$.   Thus we have the algebra 
\begin{equation}\label{3Clifford}
\{ b_j, b_k \} ~=~ 2\delta_{jk}.
\end{equation} 
The $b_j$ do not appear in their separate wire Hamiltonians, but we can expect to have interactions
\begin{equation}\label{3junctionH}
H_{\rm int.} ~=~ -i (\alpha \, b_1 b_2 + \beta \, b_2 b_3 + \gamma \, b_3 b_1 ) 
\end{equation}
which plausibly arise from normal or superconductive inter-wire electron hopping.   We assume here that the only important couplings among the wires involve the Majorana modes.  This is appropriate if the remaining modes are gapped and the interaction is weak -- for example, if we only include effects of resonant tunneling.   We shall relax this assumption in due course.  

We might expect, heuristically, that the interactions cause two Majorana degrees of freedom to pair up to form a conventional fermion degree of freedom, leaving one Majorana mode behind.  On the other hand, the algebra in Eqn.\,(\ref{3Clifford}) can be realized using Pauli $\sigma$ matrices, in the form $b_j = \sigma_j$.  In that realization, we have simply $H \, = \, \alpha \sigma_3 + \beta \, \sigma_1 + \gamma \, \sigma_2$.  But that Hamiltonian has eigenvalues $\pm \sqrt{| \alpha |^2 + |\beta|^2 + |\gamma |^2}$, with neither degeneracy nor zero mode.  In fact a similar problem arises even for ``junctions'' containing a single wire, since we could use $b_R = \sigma_1$ (and $b_L = \sigma_2$).   

The point is that the algebra of Eqn.\,(\ref{3Clifford}) is conceptually incomplete.  It does not incorporate relevant implications of electron number parity, or in other words electron number modulo two, which remains valid even in the presence of a pairing condensate.  The operator 
\begin{equation}
P ~\equiv~ (-1)^{N_e}
\end{equation}
that implements electron number parity should obey
\begin{eqnarray}
\label{Psquared} P^2 ~&=&~ 1 \\
\label{PHamiltonian} \left[ P, H_{\rm eff.} \right] ~&=&~ 0 \\
\label{Pb} \{ P, b_j \} ~&=&~ 0. 
\end{eqnarray}
Eqn.\,(\ref{Psquared}) follows directly from the motivating definition.  Eqn.\,(\ref{PHamiltonian}) reflects the fundamental constraint that electron number modulo two is conserved in the theories under consideration, and indeed under very broad -- possibly universal -- conditions.  Eqn.\,(\ref{Pb}) reflects, in the context of \cite{kitaevWire}, that the $b_j$ are linear functions of the $a_k, a_l^\dagger$, but is more general.  Indeed, it will persist under any ``dressing'' of the $b_j$ operators induced by interactions that conserve $P$.  Below we will see striking examples of this persistence.

The preceding puzzle can now be addressed. Including the algebra of electron parity operator, we take a concrete realization of operators as $b_1=\sigma_1 \otimes  I$,  $b_2 = \sigma_3 \otimes I$, $b_3 = \sigma_2 \otimes \sigma_1$ and $P=\sigma_2 \otimes \sigma_3$.  This choice represents the algebra  Eqns.\,(\ref{3Clifford}, \ref{Psquared}-\ref{Pb}). The Hamiltonian represented in this enlarged space contains doublets at each energy level.  (Related algebraic structures are implicit in \cite{akhmerov}.  See also \cite{modeTransport, modeTransport3D, modeTransportJJA, modeTransportQ1DNet} for additional constructions.)

%
%

Returning now to the abstract analysis, consider the special operator
\begin{equation}
\Gamma ~\equiv~ - i b_1 b_2 b_3.
\end{equation}
It satisfies 
\begin{eqnarray}\label{goodGammaProperties}
\label{GammaSquared} \Gamma^2 ~&=&~ 1 \\
\label{Gammab}  \left[ \Gamma, b_j \right] ~&=&~ 0 \\
\label{GammaH} \left[ \Gamma, H_{\rm eff.} \right] ~&=&~ 0 \\
\label{GammaP} \{ \Gamma, P \} ~&=&~ 0. 
\end{eqnarray}
Eqns.\,(\ref{GammaSquared}, \ref{Gammab}) follow directly from the definition, while Eqn.\,(\ref{GammaH}) follows, given Eqn.\,(\ref{Gammab}),  from the requirement that $H_{\rm eff.}$ should contain only terms of even degree in the $b_i$s.  That requirement, in turn, follows from the restriction of the Hamiltonian to terms even under $P$.   Finally Eqn.\,(\ref{GammaP}) is a direct consequence of Eqn.(\ref{Pb}) and the definition of $\Gamma$.   

This emergent $\Gamma$ has the characteristic properties of a Majorana mode operator: It is hermitean, it squares to one, and it has odd electron number parity.   Most crucially, it commutes with the Hamiltonian, but is not a function of the Hamiltonian.   We can bring the relevant structure into focus by going to a basis where $H$ and $P$ are both diagonal.   Then from Eqn.\,(\ref{GammaP}), we see that $\Gamma$ takes states with $P = \pm 1$ into states of the same energy with $P = \mp 1$.    This doubling applies to all energy eigenstates, not only to the ground state.   It is reminiscent of, but differs from, Kramers doubling.  (No antiunitary operation appears, nor is $T$ symmetry assumed.)

One also has a linear operator 
\begin{equation}\label{linearOp}
w ~\equiv~ \, \alpha \, b_3 + \beta \, b_1 + \gamma \, b_2 \, 
\end{equation}
that commutes with the Hamiltonian.  
But this $w$ is not independent of $\Gamma$, since we have
\begin{equation}
w ~=~   H \, \Gamma.
\end{equation}
and it is $\Gamma$, not $w$, which generalizes smoothly.

The same considerations apply to a junction supporting any odd number $p$ of Majorana mode operators, with 
\begin{equation}
\Gamma ~\equiv~ i^{\frac{p(p-1)}{2}} \, \prod\limits_{j=1}^p \, \gamma_j.
\end{equation}
For even $p$, however, we get a commutator instead of an anticommutator in Eqn.(\ref{GammaP}), and the doubling construction fails.   For odd $p \geq 5$ generally there is no linear operator, analogous to the $w$ of Eqn.\,(\ref{linearOp}), that commutes with $H$. (If the Hamiltonian is quadratic, the existence of a linear zero mode follows from simple linear algebra -- namely, the existence of a zero eigenvalue of an odd-dimensional antisymmetric matrix, as discussed in many earlier analyses.  But for more complex, realistic Hamiltonians, including nearby electron modes as envisaged below, that argument is insufficient, even for $p=3$. The emergent operator $\Gamma$, on the other hand, always commutes with the Hamiltonian (Eqns. (\ref{GammaH})), even allowing for higher order contributions such as quartic or higher polynomials in the $b_i$s.)

Now let us revisit the approximation of keeping only the interactions of the Majorana modes from the separate wires.  We can in fact, without difficulty, include any finite number of ``ordinary'' creation-annihilation modes from each wire, thus including all degrees of freedom that overlap significantly with the junction under consideration.    These can be analyzed, as in Eqn.\,(\ref{clifford}), into an even number of additional $\gamma$ operators, to include with the odd number of $b_j$.   But then the product $\Gamma^\prime$ of all these operators, including both types (and the appropriate power of $i$), retains the good properties Eqn.\,(\ref{goodGammaProperties}) of the $\Gamma$ operator we had before.  

If $p \geq 5$, or even at $p=3$ with nearby electron interactions included effects, the emergent zero mode is highly non-linear entangled state involving all the wires that participate at the junction. The robustness of these conclusions results from the algebraic properties of $\Gamma$ we identified. 
If we have a circuit with several junctions $j$, the emergent $\Gamma_j$ will obey the Clifford algebra 
\begin{equation}
\{ \Gamma_j, \Gamma_k \} ~=~ 2 \delta_{jk}.
\end{equation}
This applies also to junctions with $p =1$, i.e. simple terminals; nor need the circuit be connected.

The algebraic structure defined by Eqns.\,(\ref{Psquared}-\ref{Pb}) is fully non-perturbative.   It may be taken as the definition of the universality class supporting Majorana modes.  The construction of  $\Gamma$ (in its most general form) and its consequences Eqns.\,(\ref{GammaSquared}-\ref{GammaP}) reproduces that structure, allowing for additional interactions, with $\Gamma$ playing the role of an emergent $b$. The definition of $\Gamma$, the consequences Eqns.\,(\ref{GammaSquared}-\ref{GammaP}), and the deduction of doubling are likewise fully non-perturbative.  It is noteworthy that our construction of emergent Majorinos is at the opposite extreme from a single-particle operator: The mode it excites is associated with the {\it product\/} wave function over the modes associated with the $b_j$, rather than a linear combination.    In this sense we have extreme valence-bond (Heitler-London) as opposed to linear (Mulliken) orbitals.  The contrast is especially marked, of course, for large $p$. 

When there are several junctions $k$, each has its own $\Gamma_k$ operator, and together they define a Clifford algebra
\begin{equation}\label{cliffordMajorino}
\{ \Gamma_k, \Gamma_l \} ~=~ 2 \delta_{kl}
\end{equation}
as is easily demonstrated.  Thus Majorinos at different junctions anticommute, similar to fermions.   But the square of each $\Gamma_k$ is unity, not zero, so they do not obey the Pauli exclusion principle.    They are neither bosons nor fermions, but a type of nonabelian anyon, whose quantum statistics is effectively defined by Eqn.\,(\ref{cliffordMajorino}).  

The simple, explicit construction of emergent Majorinos in terms of the underlying microscopic variables might be helpful in the design of useful circuit operations, based on their response to variation of the Hamiltonian parameters, which could be controlled by applying external fields.  

\subsection{Continuum Majorinos}

We can also define a simple continuum construction, based directly on the Majorana equation, that supports Majorinos.   It is a modification of the classic Jackiw-Rebbi \cite{jackiwRebbi} model, where in place of ordinary mass the background field imparts Majorana mass to an electron, and is closely related to considerations in \cite{fu}.

To describe it in a self-contained manner, let us consider a 1+1 dimensional version of Majorana's equation Eqn.(\ref{MajoranaEquation}), allowing both for Majorana mass $M$ and conventional mass $m$
\begin{equation}
i \gamma^\mu \partial_\mu \psi + M(x) \psi^{\ *} \, + \, m \psi ~=~ 0 
\end{equation}
where we allow the Majorana mass $M(x)$ to depend on $x$.   For our $\gamma$ matrices, we take simply
\begin{eqnarray}
\gamma^0 ~&=&~ \sigma_2 \nonumber \\
\gamma^1 ~&=&~ i \sigma_3 
\end{eqnarray}
The equation for a zero-energy solution is then
\begin{equation}
\sigma_3 \frac{d\psi}{dx} ~=~ M(x) \psi^* + m \psi
\end{equation}
or, in terms of the real and imaginary parts $\psi = \phi + i \eta$,
\begin{eqnarray}
\sigma_3 \frac{d\phi}{dx} ~&=&~ \bigl( m + M(x) \bigr) \phi \nonumber \\
\sigma_3 \frac{d\eta}{dx} ~&=&~ \bigl( m - M(x)  \bigr) \eta
\end{eqnarray} 
Now writing the spinors in terms of $\sigma_3$ eigenspinors (i.e., up and down components) $\phi_\pm, \eta_\pm$
we find the formal solutions
\begin{eqnarray}\label{candidateSolutions}
\phi_\pm (x) ~&=&~ \phi_\pm (0) \exp \pm \int\limits_0^x \, dy \, \bigl( m + M(y) \bigr) \nonumber \\
\eta_\pm (x) ~&=&~ \eta_\pm (0) \exp \pm \int\limits_0^x \, dy \, \bigl( m - M(y) \bigr) 
\end{eqnarray}
These solutions will be normalizable only if the integrand is negative for large positive $y$ and positive for large negative $y$.   It is not difficult to construct monotonic profiles for $M(y)$ and values for $m$ such that exactly one of the candidate solutions is normalizable.  In that case one will have a single Majorana mode localized, after an appropriate shift, around $x=0$.


\begin{thebibliography}{99}

\bibitem{spinPaper} E. Majorana (1932).  Il Nuovo Cimento 9 (2): 43�50. doi:10.1007/BF02960953.

\bibitem{blochRabi} F.~Bloch and I.~I.~Rabi (1945).  Rev. Mod. Phys. 17, 237. 

\bibitem{stueckelberg} E.~Stueckelberg (1932).  Helvetica Physica Acta 5: 369. doi:10.5169/seals-110177.

\bibitem{landau}  L.~Landau (1932).  Physikalische Zeitschrift der Sowjetunion 2: 46�51.

\bibitem{zener} C.~Zener (1932). Proceedings of the Royal Society of London A 137 (6): 696�702. Bibcode:1932RSPSA.137..696Z. doi:10.1098/rspa.1932.0165. JSTOR 96038.


\bibitem{fermionPaper} E.~Majorana Nuovo Cimento 5, 171 (1937).

\bibitem{particleChapter} E.~Akhmedov, chapter on Majorana's influence in particle physics, this book.

\bibitem{majorinoAlgebra} J.~Lee and F.~Wilczek Phys. Rev. Lett. 111, 226402 (2013).

\bibitem{kitaevWire} {
  A.~Y.~Kitaev,
  Phys.~Usp.\  {\bf 44}, no. 10S, 131 (2001); arxiv:cond-mat/0010440.
  }

\bibitem{symmetricSpinor} L.~Landau and E.~Lifshitz, {\it Quantum Mechanics} \S 57 (Pergamon Press, 1965)

\bibitem{crossing} A.~Shapere and F.~Wilczek ``Transitions and Phases in Nonabelian Crossing'' (paper in preparation). 



\bibitem{dirac} P.~A.~M.~Dirac Proc. R. Soc. A 117, 610�624 (1928).

\bibitem{reines} C.~Cowan, F.~Reines, F.~Harrison, H.~Kruse and A.~McGuire Science 124, 103�104 (1956).


\bibitem{neutrinoOscillations} Y.~Fukuda et al. Phys. Rev. Lett. 81, 1562�1567 (1998).

\bibitem{nambu} Y.~Nambu and G.~Jona-Lasinio, Phys. Rev. 122, 345-358 (1961) doi:10.1103/PhysRev.122.345; Phys. Rev. 124, 246-254 (1961) doi: 10.1103/PhysRev.124.246

\bibitem{beenakker} C.~Beenakker, Phys.Rev.Lett. 112, 070604 (2014).

\bibitem{Mourik2012}{
V.~Mourik, K.~Zou, S.~M.~Frolov, S.~R.~Plissard, E.~P.~A.~M.~Bakkers and L.~P.~Kouwenhoven,
Science \ {\bf 336}, 1003 (2012).
}

\bibitem{JackiwRossi}{
R.~Jackiw and P.~Rossi,
Nucl.~Phys.~B \ {\bf 190}, 681 (1981).
}

\bibitem{MooreRead}{
G.~Moore and N.~Read, 
Nucl.~Phys.~B \ {\bf 360}, 362 (1991).
}

\bibitem{ReadGreen}{
N.~Read and D.~Green,
Phys.~Rev.~B \ {\bf 61}, 10267 (2000).
}

\bibitem{AliceaRev}{
J.~Alicea,
Rep.~Prog.~Phys. \ {\bf 75}, 076501 (2012).
}

\bibitem{BeenakkerRev}{
C.~W.~J.~Beenakker,
Annu.~Rev.~Con.~Mat.~Phys. \ {\bf 4}, 113 (2013)
}

\bibitem{Wilczek2009}{
F.~Wilczek,
Nature Physics \ {\bf 5}, 614 (2009).
}

\bibitem{Kitaev2006}{
  A.~Y.~Kitaev,
  A.~Ann.~Phys.\ {\bf 321}, 2 (2006).
}

\bibitem{NayakRMP}{
C.~Nayak, C.~H.~Simons, A.~Stern, M.~Freedman and S.~Das~Sarma,
Rev.~Mod.~Phys. \ {\bf 80}, 1083 (2008).
}

\bibitem{ericeWorkshop}{
{\it Conference on Majorana Physics in Condensed Matter : }  web.nano.cnr.it/mpcm13/MPCM2013\_Booklet.pdf},
and references therein.


\bibitem{akhmerov} {
 A.~R.~Akhmerov,
  Phys.~Rev.~B {\bf 82}, 020509 (2010).
  }
  
\bibitem{modeTransport}{
  J.~Alicea, Y.~Oreg, G.~Refael, F.~von Oppen, and M.~P. ~A.~ Fisher,
  Nature Physics {\bf 7}, 412 (2010).
  }
  
\bibitem{modeTransport3D}{
B.~I.~Halperin, Y.~Oreg, A.~Stern, G.~Refael, J.~Alicea and F.~von~Oppen,
Phys.~Rev.~B {\bf 85}, 144501 (2012). 
}

\bibitem{modeTransportJJA}{
B.~van~Heck, A.~R.~Arkmerov, F.~Hassler, M.~Burrello and C.~W.~J.~Beenakker,
New J.~Phys. {\bf 14}, 035019 (2012).
}

\bibitem{modeTransportQ1DNet}{
D.~J.~Clarke, J.~D.~Sau, S.~Tewari,
Phys.~Rev.~B {\bf 84}, 035120 (2011).
}

\bibitem{jackiwRebbi} R.~Jackiw and C.~Rebbi Phys. Rev. D {\bf 13} 3398 (1976).


\bibitem{fu} L.~Fu and C.~Kane Phys. Rev. B {\bf 79}, 161408(R) (2009).
  

\end{thebibliography}
\end{document}